\documentclass[conference, 10 pt]{IEEEtran}

\usepackage{soul} 
\IEEEoverridecommandlockouts
\usepackage{cite}
\usepackage{amsmath,amssymb,amsfonts}
\usepackage{braket}
\usepackage{algorithmic}
\usepackage{multirow}
\usepackage{graphicx}
\usepackage{textcomp}
\usepackage{xcolor}
\def\BibTeX{{\rm B\kern-.05em{\sc i\kern-.025em b}\kern-.08em
    T\kern-.1667em\lower.7ex\hbox{E}\kern-.125emX}}
\begin{document}

\title{Trojan Taxonomy in Quantum Computing\\
}

\author{\IEEEauthorblockN{Subrata Das}
\IEEEauthorblockA{\textit{Dept. of Electrical Engineering} \\
\textit{Pennsylvania State University}\\
University Park, PA \\
sjd6366@psu.edu}
\and
\IEEEauthorblockN{Swaroop Ghosh}
\IEEEauthorblockA{\textit{Dept. of Electrical Engineering} \\
\textit{Pennsylvania State University}\\
University Park, PA \\
szg212@psu.edu}

}

\maketitle

\begin{abstract}
Quantum computing introduces unfamiliar security vulnerabilities demanding customized threat models. Hardware and software Trojans pose serious concerns needing rethinking from classical paradigms. This paper develops the first structured taxonomy of Trojans tailored to quantum information systems. We enumerate potential attack vectors across the quantum stack from hardware to software layers. A categorization of quantum Trojan types and payloads is outlined ranging from reliability degradation, functionality corruption, backdoors, and denial-of-service. Adversarial motivations behind quantum Trojans are analyzed. By consolidating diverse threats into a unified perspective, this quantum Trojan taxonomy provides insights guiding threat modeling, risk analysis, detection mechanisms, and security best practices customized for this novel computing paradigm.

\end{abstract}


\section{Introduction}
\label{sec:intro}

Quantum computing has emerged as a revolutionary technology with the potential to solve complex problems that are intractable for classical computers. The advent of quantum computing promises significant advancements in various fields, such as cryptography, optimization, simulation, and machine learning. As quantum computers advance toward practical applications, ensuring their secure and reliable operation is imperative but also presents unfamiliar challenges. While quantum technologies promise advantages over classical systems, they also introduce poorly understood security vulnerabilities that classical threat models cannot adequately address. This necessitates developing new perspectives on hardware and software security intrinsically tailored to quantum information systems. In particular, hardware Trojans pose a serious concern for quantum computing that demands rethinking. In classical domain, hardware Trojans are malicious modifications to a system's physical implementation, such as tampering with integrated circuits during fabrication \cite{tehranipoor2011introduction}. They aim to alter functionality or leak sensitive data through inserted backdoors. Software Trojans similarly inject malicious code into programs and applications to distort behavior or outcomes \cite{haagman2005trojan}. Extensive research has explored hardware Trojan taxonomies, detection techniques, and defenses for classical systems \cite{bhunia2014hardware, xiao2016hardware}. Classical hardware Trojan taxonomy categorizes threats based on insertion phase (design or fabrication), abstraction level (transistor, gate, register transfer level), activation mechanisms (always on, externally triggered), and impacts (denial-of-service, reducing reliability, leaking secrets) \cite{bhunia2014hardware}. Classical software Trojan taxonomies characterize threats based on infection techniques, triggering mechanisms, and payloads \cite{haagman2005trojan}. 

However, translating these classical Trojan models directly to quantum computing can overlook unique risks and opportunities. For instance, in conventional digital circuits, hardware Trojans can be detected post-manufacturing by applying test patterns and validating the outputs against expected patterns \cite{trusthub}. 
However, testing-based approaches are ineffective for quantum circuits as the user lacks an oracle to verify results. 
Quantum-specific characteristics necessitate developing custom Trojan models from the ground up that account for superposition, entanglement, qubit measurement, and qubit operations. Rethinking Trojans for quantum algorithms, software tools, hardware devices, and quantum data is crucial to develop defense mechanisms. A taxonomy of quantum Trojans can guide threat modeling, risk analysis, detection mechanisms, and defenses tailored to this unique computing paradigm.


\textbf{Contributions:}
We present a taxonomy that categorizes potential Trojan attacks across the quantum technology stack. The taxonomy is structured into four key dimensions: insertion points, Trojan types, payloads, and adversarial objectives. First, we analyze insertion points for Trojans across quantum hardware manufacturing, software and compilers, algorithm design, data generation, and network transmission layers. Next, we enumerate common quantum Trojan types. We also outline likely payloads and computational impacts for each Trojan type. Finally, we summarize plausible adversarial objectives. This multi-dimensional taxonomy provides a customized systematic characterization of Trojan threats tailored to the unique features of quantum information systems. By enumerating potential dangers early, the proposed taxonomy aims to guide threat modeling, risk analysis, detection mechanisms, and security best practices as quantum computers advance toward real-world deployment.





\section{Classical vs. Quantum Trojans}

Classical computing has a long history of research into hardware Trojans, software vulnerabilities, and side-channel attacks. Techniques have been developed for detecting, preventing, and mitigating such threats in conventional digital logic systems. However, quantum computing introduces unfamiliar and dimly understood risks that demand rethinking Trojan models from the ground up. Quantum computers encode information in qubits as probabilistic superposition states. Computation occurs by applying sequences of fundamental quantum logic gates that leverage principles like superposition, entanglement, and measurement. This quantum mechanical approach to processing information leads to fundamentally different vulnerabilities compared to deterministic classical digital logic. Several intrinsic characteristics of quantum technologies severely limit the applicability of traditional Trojan detection techniques, which will be discussed in this section.

\textit{\textbf{Absence of a Verification Oracle:}} In conventional digital circuits, hardware Trojans can be detected post-manufacturing by running test patterns and checking if the circuit outputs match the expected values from a golden model. Test vectors can be cleverly crafted to trigger rare Trojan
activation, revealing incorrect outputs that imply the presence of Trojan. Conversely, the Trojan should be designed to bypass such detection methodologies. However, this testing methodology fails for quantum processors since users lack a verification oracle. This is true since practical quantum circuits cannot be simulated in classical computers in a reasonable time and therefore, the output of the circuit is not known. This also eases the Trojan design process since it does not need to be activated rarely anymore. 
The inability to validate quantum circuits with an oracle significantly hinders classical Trojan detection approaches. 

\textbf{\textit{Intertwined Triggers and Payloads:}} Classical hardware Trojans rely on rare trigger conditions to activate malicious payload circuits after deployment. During functional validation and testing, the triggers stay inactive to avoid detection. Adversaries exploit this gap between design-time testing and operation in the field. 
In quantum computation, the expected outputs are not known a priori for users to check for errors. This renders the concept of rarely triggered payloads less meaningful. Moreover, quantum properties like entanglement allow crafting Trojans where the effects are intrinsically linked to the trigger conditions. For example, a multi-qubit gate could be designed to intentionally reduce entanglement fidelity only when operating on certain qubit combinations. The gate's weakened operation triggers errors precisely when applied during computation on specific inputs. Similarly, measurement operations could be designed to selectively return incorrect results only for particular superposition states. The act of observing the system triggers the payload. This intertwining of trigger and payload is facilitated by quantum superposition, entanglement, and measurement. The payload effects manifest precisely when the trigger circuitry is activated by the input state. This blurs the line between trigger and payload, which is central to classical Trojan architectures.

\textbf{\textit{Limited Observable Side-Channels:}} In classical hardware Trojan detection, side-channel analysis techniques like measuring power consumption, timing delays, or electromagnetic radiation are commonly used. The presence of a Trojan circuit often introduces subtle anomalies detectable in these physical side-channels. However, when accessing quantum computers in a cloud service model, users have very restricted visibility into the physical hardware implementing the qubits. The quantum processor's power consumption, thermal profiles, and electromagnetic signatures are not available to the users. Any additional Trojan logic inserted into the quantum circuit would invariably increase the power dissipation, since the chip needs to be cooled to millikelvin temperatures. Extra gates cause more heat generation requiring higher cryogenic cooling power, which could act as a side-channel signature. However, without physical access this information is obscured from users. Similarly, Trojans may increase overall execution time of algorithms by requiring more operations. But users do not have an expected timeline to compare against to detect potential execution delays. There is also no concept of chip "area" overhead in gate-model quantum computing since physical layout is not exposed. Moreover, the intrinsic uncertainties and noise in superposition, entanglement, and measurement may easily overshadow any subtle side-channel anomalies caused by Trojans. With current NISQ processors, decoherence and gate errors are dominant effects compared to the marginal overhead of stealthy Trojans. This further limits the availability of reliable side-channel signatures that could be leveraged for Trojan detection by users of cloud quantum services.

\textbf{\textit{Focus on Parameter Tampering:}} While classical Trojans aim to introduce faults into digital logic or flip bit values, quantum Trojans open up entirely new avenues of analog parameter tampering. The operation of quantum computers relies heavily on precise tuning of analog parameters. Qubits are sensitive physical systems requiring careful calibration - for example, superconducting qubits depend on microwave pulses applied at exact resonance frequencies. The fidelity of quantum gates is also closely tied to the shape and timing of control pulses. These analog settings present new targets for quantum Trojans. For instance, an adversary could manipulate the resonant frequency of a qubit by altering the bias currents or voltages applied to tune its operating point. Even a slight detuning of a few MHz could prevent the qubit from interacting with subsequent control pulses, effectively silencing it from computation. Or the coupling strengths between qubits could be weakened by changing the geometries or materials used for interconnects during fabrication. This would reduce the entanglement fidelity for multi-qubit gates. The pulse shapes used to implement quantum gates could also be tampered. Introducing pulse distortion or calibration errors in the classical electronics that generate the microwave controls would corrupt the intended gate operations. For example, a Hadamard gate pulse could be truncated early to undermine the intended superposition mapping.
Unlike classical logic faults, these types of analog parameter tampering allow continuously skewing the computation in subtle ways difficult to detect. The probabilistic nature of qubits obscured by noise provides ample cover for malicious parameter manipulation. 

\begin{figure*}[t]
    \includegraphics[width=\textwidth]{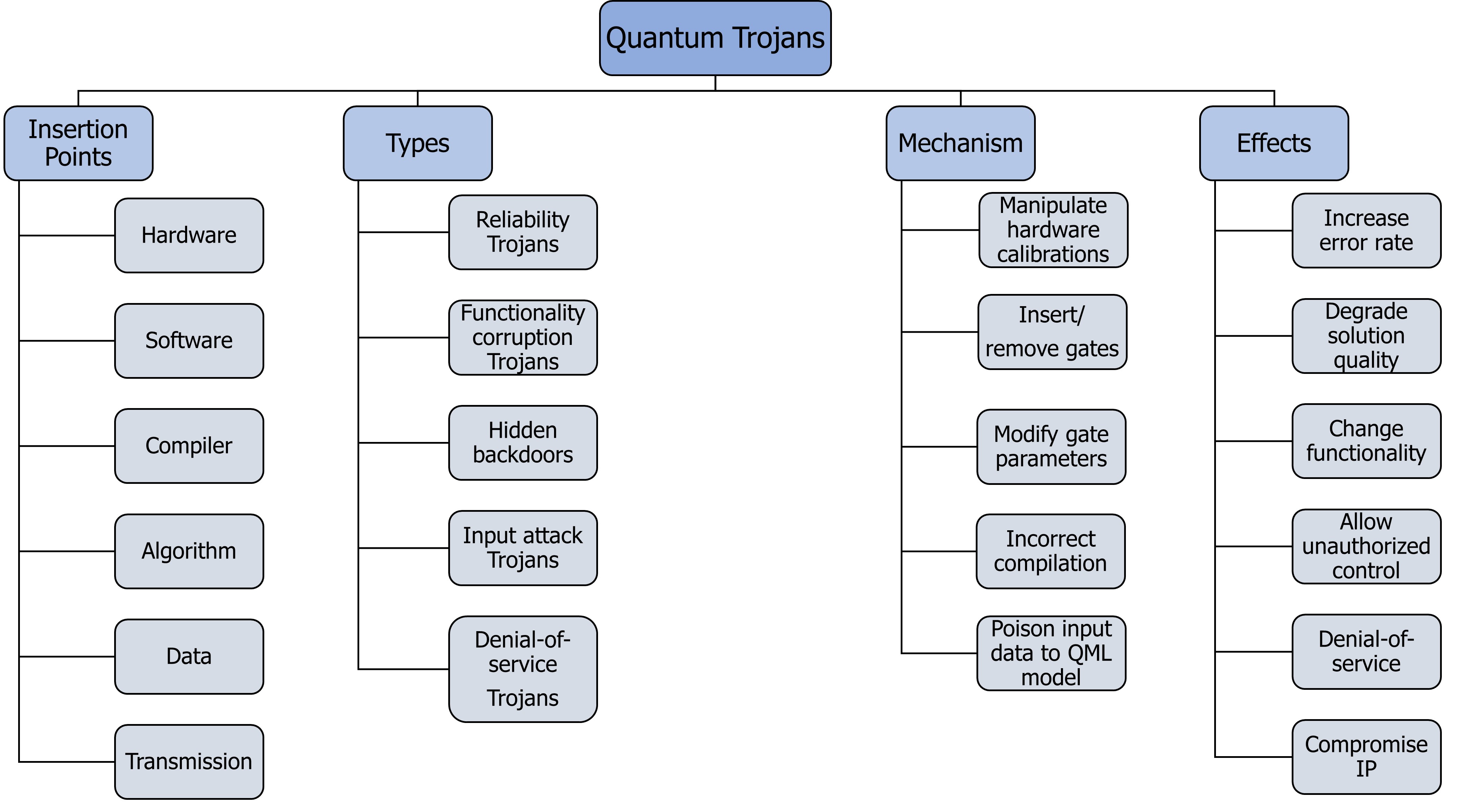}
    \centering
\caption{A comprehensive quantum Trojan taxonomy.}
    \label{fig:taxonomy}
\end{figure*}

\textbf{\textit{Extensive reliance on third parties:}} The extensive reliance on third parties in the quantum computing supply chain significantly increases the attack surface for Trojan insertion compared to classical systems. Unlike classical computers produced in mass volume, each quantum computer is essentially a specialized custom system relying on numerous external suppliers for critical components and services. For example, a user developing quantum algorithms depends on third-party compiler software to translate the high-level description into low-level quantum gates. An untrusted compiler could secretly insert Trojans by manipulating the gate decomposition or scheduling in subtle ways to introduce errors. The user has no choice but to trust the compiler's integrity. Similarly, for superconducting quantum computers, users rely on third parties for manufacturing key elements like qubit chips, cryogenic refrigerators, and control electronics tailored for each system. Insertion of Trojans directly into such hardware components by an untrusted vendor would be extremely difficult to detect yet could allow disruption or information leakage. The extensive calibration and testing services needed to characterize each qubit also require trusting third-party providers. Falsified calibration data containing Trojans could degrade the mapping of algorithms onto the qubit topology. Unlike classical systems, it is infeasible for users to fully validate or verify third-party contributions themselves due to the complexity and customization involved. This extensive trust and integration of unverified third-party software, firmware, and hardware substantially widen the attack surface available for Trojan insertion compared to classical systems where users have greater visibility into internally developed components.

\section{Taxonomy of Quantum Trojans} 
\label{sec:Taxonomy}

In this section, we develop a comprehensive taxonomy to characterize potential Trojan threats tailored to quantum information systems (Fig. \ref{fig:taxonomy}).

\subsection{Quantum Trojan Insertion Points}

\begin{figure*}[t]
    \includegraphics[width=\textwidth]{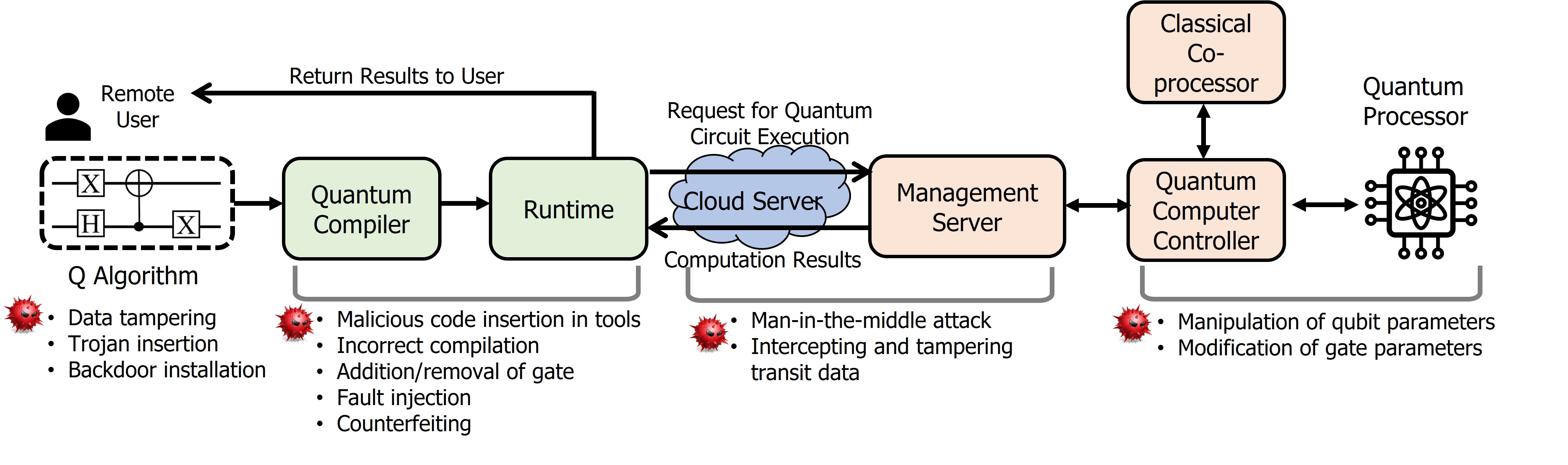}
    \centering
\caption{Trojan insertion points into the quantum computer stack.}
    \label{fig:insertion}
\end{figure*}

Quantum computing systems involve multiple entities in the supply chain during the process of designing, compiling, and running quantum circuits. These parties range from hardware manufacturers building the qubit devices to algorithm designers creating quantum circuits, to users providing input data. If any component along this supply chain becomes compromised, it opens up the possibility for Trojan insertion into the final quantum circuits. We categorize the potential compromised supply chain stages which could enable Trojan insertion into the quantum computer stack (Fig. \ref{fig:insertion}).

\textbf{\textit{Hardware:}} The hardware forms the foundation on which quantum circuits are executed. Qubit technologies used in current noisy intermediate-scale quantum (NISQ) systems include superconducting circuits, trapped ions, photonics etc \cite{das2023first}. Each qubit must be precisely tuned and calibrated, then interconnected into a larger architecture. The infrastructure to control, read out, and maintain qubit states requires extensive classical electronics and software. If the facilities and processes to produce quantum hardware become corrupted, it enables direct manipulation of qubit parameters to introduce Trojans into devices. For example, the titanium nitride or aluminum fabrication steps for superconducting qubits could be tampered to create shorted or disconnected regions. Or ion trap electrode geometries could be altered to shift qubit resonance frequencies. These malicious parameter modifications would create faults or distort computation once circuits run on the compromised hardware.

\textbf{\textit{Software and Tools:}} After the quantum hardware, the next layer in the stack is the control software and tools (e.g., compilers) used to coordinate computation \cite{ghosh2023primer}. Code libraries and drivers translate circuits into analog signals and qubit control pulses. Compilers map abstract circuits down to the specific gates executable on the given hardware topology and calibration. If adversaries infiltrate the teams designing compilers and control software, they could insert malicious code into these tools. For instance, Trojans could be embedded into compiler optimizations or pulse sequence libraries to mistranslate certain gates or algorithms. The contaminated software would then distort the output once users apply it to their own circuits. Another risk associated with relying on unreliable compilers is the potential for tampering, specifically the insertion of Trojan gates \cite{saki2021split}. When the user receives back these compiled Trojan circuits, they have no means to validate the behavior or outputs. Running the circuits on actual quantum hardware also does not reveal Trojans, since the user is unaware of the Trojan-free outputs.

\textbf{\textit{Algorithm:}}
Quantum algorithms form the next layer up the stack from low-level software. Algorithm design involves creating circuit constructions to implement mathematical formulas or computational techniques. As quantum computing gains adoption, organizations are increasingly looking to purchase or license quantum algorithms and circuit intellectual property (IP) from specialist design firms. If the vendors providing third-party quantum algorithms or circuits become untrusted, it enables the insertion of Trojans into these higher-level modules. For example, a search algorithm's oracle could be planted with trapdoors to selectively introduce errors into certain problem instances or a quantum chemistry circuit sold as an IP could have certain unitary blocks designed to distort energy calculations. These backdoors would propagate errors once the compromised circuits are integrated into users' applications.

\textbf{\textit{Data:}}
The final supply chain source of potential Trojans is at the data level. Many quantum algorithms take real-world data as inputs, such as financial transactions or molecular configurations. If this input data becomes corrupted, it could induce errors in the computational results. Data poisoning is a well-known classical machine-learning attack vector but has not been explored much in quantum contexts. For instance, tampered training data in a quantum classifier circuit could introduce mislabeled samples that then skew decision boundaries after training. Or precision issues in molecular data could shift simulations of chemical properties. Any data dependencies should be scrutinized as potential Trojan insertion points.

\textbf{\textit{Transmission:}} Another potential quantum Trojan insertion point is during the transmission of quantum circuits from the user's computer to the cloud servers. Adversaries could attempt man-in-the-middle (MITM) attacks by intercepting the circuit data in transit and tampering with it before forwarding it to the quantum hardware. For example, a malicious party could intercept the network request containing the quantum circuit and modify or insert additional gates into the code before sending it to the quantum cloud service for execution. This attack would rely on the circuits being transmitted without encryption or two-level attack where adversary will first compromise the encryption of the victim's classical computer followed by MITM attack. 

\subsection{Types of Quantum Trojans and Impacts}

The types of Quantum Trojans adversaries can insert depend on their goals and the layer being targeted. Here, we outline common categories tailored to quantum systems and plausible impacts for each Trojan type.

\textbf{\textit{Reliability Degradation:}} These aim to degrade the performance and fidelity of quantum computations by manipulating qubit parameters or introducing defects. Examples include altering quantum gate calibrations to skew fidelities, modifying qubit frequencies or coupling strengths, or weakening entanglement operations. The resulting loss in reliability manifests as increased error rates or reduced success probabilities once algorithms execute on the compromised system. By reducing success probabilities, these Trojans can force the user to expend more quantum resources like the number of shots to extract correct solutions. For example, Trojans could increase the required number of measurements or circuit repetitions needed to reliably read out results. This amplifies the time and financial costs of running quantum programs. Another core impact of reliability Trojans is the degraded quality of the solutions found by quantum algorithms. For instance, a Quantum Approximate Optimization Algorithm (QAOA) circuit for portfolio optimization may originally yield 95\% optimal allocations, but with a Trojan, this could drop to 60\%.

\textbf{\textit{Functionality Corruption:}} These directly distort the core functionality or outputs of quantum algorithms by tampering with circuit operations. For instance, adversaries could incorrectly compile certain gates or subroutines, insert extra gates, or tweak gate parameters. The Trojan alters the algorithm itself rather than just reducing reliability. This leads to outright incorrect solutions or biases that skew certain problem outcomes. For instance, an adversarial quantum machine learning (QML) model could be trained to misclassify certain samples by design. Or a simulation's predictions could be intentionally shifted for particular molecular configurations.

\textbf{\textit{Hidden Backdoors:}} Hidden backdoors allow adversaries secret access or control over quantum computations through planted vulnerabilities \cite{chu2023qtrojan}. For example, an adversary could insert a trapdoor into a quantum cryptanalysis algorithm's collision-finding routine. This would enable them to selectively fail the routine for certain inputs chosen by the attacker, compromising security. Another approach is embedding undocumented parameters or switches into QML models or simulators. The adversary could use these to subtly tweak the training process or distort outputs for targeted samples without the user's knowledge. Backdoors could also take the form of hardcoded credentials planted into control software, granting adversaries access to quantum infrastructure. Or secret built-in kill switches that shut down quantum processors when activated by the attacker. 

\textit{\textbf{Input Manipulation:}} While not representing full Trojans inserted into the quantum stack, adversaries could also craft malicious inputs designed to induce errors or skew computations \cite{huang2019performance}. These input attacks leverage algorithms' data dependencies as attack vectors. For instance, an adversary could poison the training data used for a quantum machine learning classifier. By intentionally corrupting labels in the training set, the adversary could degrade the model's test accuracy once trained on this data. For example, some false labels could be flipped to incorrectly indicate malicious software executables as benign files. The classifier trained on this tampered data would have skewed decision boundaries, causing targeted mispredictions desired by the attacker. Another example is tweaking molecular configurations fed into quantum chemistry simulations. An adversary could make precision alterations to bond lengths or angles that shift the predicted energy levels to invalidate the physics modeling. Similar input data poisoning could distort solutions for quantum algorithms in optimization, finance, or other domains.

\textit{\textbf{Denial-of-Service:}} A particularly damaging Trojan payload could be completely denying execution of users' quantum programs on the hardware. This constitutes true denial-of-service for quantum computing. Rather than just introducing functional corruption or degrading reliability, a denial-of-service Trojan could halt computation entirely. For example, it could crash the classical servers and electronics controlling the quantum processor, preventing operation. Or trigger conditions that damage the quantum chip to make it unusable. This mirrors classical denial-of-service attacks that overload systems and make services inaccessible to legitimate users. However, translating such aggressive attacks to quantum computing is challenging. One potential approach could be crafting malicious quantum circuits that intentionally overload and disrupt the classical hardware interface that controls qubit operations. This interface decodes instructions, generates microwave pulses, measures outputs, and processes results. By intently inducing a crash through malformed programs, adversaries could shut out users. However, creating robust quantum programs that reliably induce such software crashes is difficult with current knowledge. Further research into the classical-quantum interface and prototyping denial-of-service code could reveal additional attack vectors.

\section{Adversarial Objectives of Quantum Trojans}

This section summarizes common adversarial motivations for inserting Trojans across the quantum computing stack. Understanding these goals provides insights into likely Trojan designs, payloads, and targets. Awareness of objectives guides the development of threat models, risk assessments, and countermeasures against quantum Trojans.

\textbf{\textit{Compromising Intellectual Property (IP):}} A major incentive behind inserting Trojans into third-party quantum circuits or software is gaining unauthorized access to proprietary IP. For instance, an untrusted quantum chemistry simulation vendor could plant Trojans that intentionally leak a pharmaceutical company's confidential molecular models during computation. Or a search algorithm provider may embed backdoors to extract database contents being queried. By compromising IP, attackers can steal sensitive data assets and reverse engineer details of a victim's R\&D.

\textbf{\textit{Manipulating Computational Outcomes:}} Trojans give adversaries a pathway to intentionally manipulate the outcomes of quantum algorithms for financial, political, or other gains. For example, attackers could tweak a quantum trading algorithm to bias investment recommendations and profit off market movements. Or Trojans planted in quantum machine learning classifiers may induce targeted mispredictions that benefit the adversary's interests. Unlike reliability failures, these manipulation Trojans deterministically and discreetly skew outputs.

\textbf{\textit{Denying Quantum-as-a-Service (QCaaS) Access:}} With quantum computing transitioning to a cloud service model, denial-of-service Trojans pose a threat to service availability. By crashing server infrastructure or even damaging quantum chips, adversaries could completely prevent users from harnessing quantum computing resources. This forces dependency on only compromised Quantum-as-a-Service (QCaaS) vendors. Constructing robust denial-of-service Trojans is non-trivial, but still poses theoretical risks.

\textbf{\textit{Degrading Computational Value:}} Trojans that reduce solution quality, lower success probabilities, or increase costs diminish the practical value quantum computers offer users. For instance, optimization problems may only reach 60\% optimality compared to 95\% on Trojan-free hardware. This degradation incentivizes sticking with classical methods and slows mainstream quantum adoption. Subtly reducing computational value allows adversaries to deter quantum computational growth.

\textbf{\textit{Cloaking QC Vulnerabilities}} In rare cases, adversaries may even use Trojans to cloak existing flaws in quantum processors and prevent further scrutiny. For example, reliability Trojans could worsen gate fidelities and amplify a hardware manufacturer's fabrication defects. With computation failing more frequently, users may not detect the underlying vulnerabilities being masked. However, easier methods like input attacks likely serve better for cloaking.

\section{Conclusion}
\label{sec:conclusion}

This paper presented the first taxonomy consolidatedly characterizing Trojan threats customized for quantum computing across hardware, software, algorithms, and data. A multi-dimensional taxonomy analyzing quantum Trojan insertion points, types, payloads, and attack goals has been proposed. As quantum computers continue maturing toward real-world applications, ensuring their reliable and secure operation will be imperative. The proposed quantum Trojan taxonomy provides an essential first step toward this goal by laying out a customized methodology for assessing threats in this novel and unfamiliar computing paradigm.

\section*{Acknowledgements}
This work is supported in parts by NSF (CNS-1722557, CNS-2129675, CCF-2210963, CCF-1718474, OIA-2040667, DGE-1723687, DGE-1821766, and DGE-2113839), Intel’s gift and seed grants from Penn State ICDS and Huck Institute of the Life Sciences. 

\bibliography{Manuscript}

\end{document}